# Hopping Transfer Optimizes Avalanche Multiplication in Molybdenum Disulfide

Xiaofan Cai[†], Ruichang Chen[†], Xu Gao[†], Meili Yuan[†], Haixia Hu[†], Hang Yin[†], Yuanyuan Qu[*,†], Yang Tan[*,†] and Feng Chen[*,†]

[†]School of Physics, State Key Laboratory of Crystal Materials, Shandong University, Shandong, Jinan, 250100, China,

E-mail: tanyang@sdu.edu.cn; drfchen@sdu.edu.cn; quyuanyuan@sdu.edu.cn

**ABSTRACT**

Recently, avalanche multiplication has been observed in TMDC-based FETs, enhancing sensor performance with high sensitivity. However, the high voltage required for operation can damage the FETs, making it crucial to reduce the breakdown voltage for effective sensing applications. Here, we demonstrate that the utilization of hopping transfer induced by high-density defects can effectively reduce the breakdown voltage in TMDCs FETs. By substituting oxygen atoms for sulfur atoms in a monolayer of $MoS_2$, we create $MoS_{2-x}O_x$, with $x$ carefully adjusted within the range of 0 to 0.51. Oxygen doping reduces the bandgap of TMDCs and enhances ion collision rates. Moreover, higher levels of oxygen doping ($x > 0.41$) in $MoS_{2-x}O_x$ exhibit nearest-neighbor hopping behavior, leading to a significant enhancement in electron mobility. These improvements result in a decrease in the breakdown voltage of avalanche multiplication from 26.2 V to 12.6 V. Additionally, we propose avalanche multiplication in $MoS_{2-x}O_x$ as an efficient sensing mechanism to overcome the limitations of gas sensing. The $MoS_{2-x}O_x$ sensors display an ultra-high response to $NO_2$ gas in the air, with a response of $5.8 \times 10^3$ % to $NO_2$ gas of 50 ppb at room temperature, which is nearly two orders of magnitude higher than resistance-type gas detectors based on TMDCs. This work demonstrates that hopping transfer induced by high-density oxygen defects can effectively decrease the breakdown voltage of $MoS_{2-x}O_x$ FETs, enhancing avalanche multiplication and serving as a promising mechanism for ultrasensitive gas detection.

**INTRODUCTION**

Transition metal dichalcogenides (TMDCs) have gained considerable attention as sensing materials for photo or gas detectors [1-4]. This is because of their unique structural characteristics, such as high specific surface area and strong surface adsorption capacity, as well as their excellent electronic properties, including high carrier mobility, good electrical stability, and tunable band gap [5-6]. Recently, the phenomenon of avalanche multiplication has been observed in TMDCs like $MoS_2$ and other two-dimensional TMDCs, attributing to the strong Coulomb interaction-induced quantum confinement effects [8-13]. Avalanche multiplication refers to a chain reaction process where charge carriers are accelerated by an external high voltage, leading to ionization of atoms through collisions. This generates additional carriers (electrons or holes) from the valence band [14-17]. By operating TMDC-based field-effect transistors (FETs) in an avalanche multiplication regime via gate-voltage modulation, weak signals induced by illumination or molecule adsorption can be amplified. The amplified signal through avalanche multiplication exhibits minimized background noise and ultrahigh sensitivity, enabling detectors to achieve exceptional performance [14,18]. There is a growing motivation to optimize and utilize carrier multiplication effects for photo/gas detection in order to achieve outstanding sensing performance.

The breakdown voltage ($V_b$) serves as a crucial parameter in avalanche multiplication-based detectors, which refers to the phenomenon of generating additional carriers through impact ionization [14,19-21]. It is highly desirable to minimize $V_b$, as excessively high values can lead to damage in TMDC-based FETs during repeated detection measurements. Extensive research has been carried out to investigate the influence of carrier type, TMDCs thickness, and length on avalanche multiplication [8,9,15,16]. It is currently believed that enhancing the collision probability of carriers per unit length ($α$) and reducing the TMDCs bandgap are effective strategies to reduce $V_b$ [8,14]. Achieving these objectives require overcoming the inherent limitations of pristine materials. In this context, the customized design and engineering of defects offer a highly promising approach.

Defects, such as vacancies and impurities, greatly affect the bandgap of TMDCs and can also cause carrier scattering due to random potential fluctuations [22-26]. However, defects in TMDCs also lead to carrier trapping and reduction of carrier mobility, hindering avalanche multiplication [27-30]. To address this contradiction, we propose the concept of hopping transfer between nearest-neighbor defects. Nearest-neighbor hopping (NNH) is a common phenomenon in polymers and materials with high defect concentrations, allowing for the generation of high carrier concentrations in defective materials [31-34]. To tackle this challenge, our research focuses on avalanche multiplication in $MoS_2$ with high defect concentrations.

In this work, we present the impact of NNH on the avalanche multiplication performance of $MoS_2$. By introducing oxygen ($O$) atoms to replace sulfur ($S$) atoms in a monolayer of $MoS_2$, we create defective structures known as $MoS_{2-x}O_x$, with $x$ carefully controlled within the range of 0 to 0.51. When $x$ exceeds 0.44, heavily doped $O$ defects enable $MoS_{2-x}O_x$ to exhibit NNH. Compared to pristine $MoS_2$, the presence of $O$ defects in $MoS_{2-x}O_x$ increases the collision probability of carriers, while maintaining a carrier mobility comparable to that of $MoS_2$. Remarkably, the NNH effect in $MoS_{2-x}O_x$ leads to a higher susceptibility to avalanche multiplication compared to $MoS_2$, with a significant reduction in the breakdown voltage ($V_b$) from 26.2 V to 12.6 V under the same measurement conditions. Furthermore, we have developed a gas sensor utilizing the avalanche multiplication characteristics of $MoS_{2-x}O_x$. This avalanche multiplication sensor ($MoS_{2-0.51}O_{0.51}$) has a LOD of approximately $1.4\times10^{-4}$ ppb and exhibits exceptional gas response to 50 ppb $NO_2$ gas at room temperature, with a remarkable signal responsivity of $5.8\times10^3$%, surpassing traditional resistance-type gas detectors based on TMDCs by two orders of magnitude. Our $MoS_{2-x}O_x$ gas sensor FETs operate based on the properties of avalanche multiplication, unlike conventional breakdown-type detectors that employ two parallel electrodes for gas ionization and breakdown voltage measurement [35-39].

Our sensor is versatile and suitable for gas concentration detection, particularly for low concentrations, as it does not require specific gas pressures. This work demonstrates the modulation of avalanche multiplication characteristics through hopping transfer and highlights avalanche multiplication as an efficient physical mechanism for the development of ultrasensitive gas sensors.

**Preparation of MoS$_{2-x}$O$_x$ FET**

The monolayer MoS$_2$ film used in this study is synthesized using the chemical vapor deposition (CVD) technique. Supplementary I displays detailed characteristics of the MoS$_2$ monolayer. We transfer the MoS$_2$ film onto a pair of golden electrodes (*D* and *S*) with a separation distance of 10 μm (Fig. 1a). These electrodes are placed on a SiO$_2$/Si substrate, where the SiO$_2$ layer has a thickness of 300 nm, and the Si substrate is heavily doped with $P^{++}$ silicon. The contact quality of the MoS$_2$ film and electrodes is determined by the output curves ($I_{DS}$-$V_{DS}$) in Supplementary II. The forward $I_{DS}$-$V_{DS}$ curve is displayed in a logarithmic scale and fitted following the equation $I_{DS}=C_0 V_{DS}^{\gamma}$, where $C_0$ is the a constant and $\gamma$ is the parameter to qualify the average linearity of $I_{DS}$-$V_{DS}$ curves [14]. In this situation, the $\gamma$ is determined to be 0.90 ± 0.16, demonstrating the non-perfect ohmic contact, as there are van der Waals gaps between MoS$_2$ film and Au electrodes generating the contact barrier.

*O* atoms are doped into the MoS$_2$ through a two-step process involving focused ion beam (FIB) irradiation and subsequent immersion in NO$_2$ gas (Fig. 1c). Initially, the focused *Ga$^+$* ion beam is utilized to sputter *S* atoms away from the outermost atomic layer of MoS$_2$ film, creating *S* vacancies (Fig. 1c(ii)) [40]. The ion beam energy is set at 30 keV, with varying beam currents of 7.7 pA, 24 pA, and 40 pA, respectively. To prevent any damage to the Mo atomic layer and the formation of holes in the MoS$_2$ structure, the irradiation time is carefully controlled to 500 ms. Subsequently, the irradiated film is immersed in NO$_2$ gas at the room temperature for 24 hours, allowing the *Mo* dangling bonds created by the irradiation to react fully with NO$_2$ (Fig.

1c(iii)). This reaction results in the substitution of *O* atoms for *S* vacancies, leading to the *O* doping of the irradiated film labeled as MoS$_{2-x}$O$_x$ (Fig. 1c(iv)). To ensure that the FIB irradiation process not affect the contact between the MoS$_{2-x}$O$_x$ film and gold electrodes, the irradiation area is limited to the region between the electrodes (Fig. 1b).

We determine the atomic ratio of the film after *O*-dopping treatments using XPS (X-ray photoelectron spectroscopy) spectra (Fig. 1d-f) [41,42]. After the *O*-dopping treatment, the elemental composition and atomic ratio of the sample undergo changes. The *S* atomic content gradually decreases with an increase in beam current, indicating that more *S* atoms are sputtered from the sample during FIB irradiation along with the enhanced beam current. Additionally, the peak at 236 eV gradually intensifies, corresponding to the bonding between *Mo* and *O*, demonstrating the successful incorporation of oxygen through the aforementioned process [41]. The measured atomic ratios (x) of *O*:*Mo* is 0.19(MoS$_{2-0.19}$O$_{0.19}$), 0.44 (MoS$_{2-0.44}$O$_{0.44}$) and 0.51 (MoS$_{2-0.51}$O$_{0.51}$), respectively, corresponding to beam currents of 7.7 pA, 24 pA, and 40 pA.

**High conduction Hopping behavior in MoS$_{2-x}$O$_x$ FET**

Fig. 2a illustrates the transfer characteristics of MoS$_2$ and MoS$_{2-x}$O$_x$ (x = 0.19, 0.44 and 0.51) at room temperature, presented in the logarithmic scale. The pristine MoS$_2$ exhibits the typical n-type behavior, with the on/off ratio ($V_{DS}$ = ± 20 V) of 3.7 × 10$^5$. The field-effect mobility value ($\mu$) for electrons is determined to be 10.74 cm$^2$ V$^{-1}$ s$^{-1}$ by the equation $\mu = \frac{\partial I_{ds}}{\partial V_g} \frac{L}{WC_g V_{ds}}$, where $L$ is the channel length; $W$ is the channel width; $C_g$ is the capacitance per unit area. Upon *O* doping, the on/off ratio and mobility of MoS$_{2-x}$O$_x$ vary with $x$. Fig. 2b summarizes changes in mobility of MoS$_{2-x}$O$_x$ corresponding to $x$. For $x$ = 0.19, MoS$_{2-0.19}$O$_{0.19}$ exhibits a reduced mobility of 0.02 cm$^2$ V$^{-1}$ s$^{-1}$ with a decrease of nearly three orders of magnitude compared to MoS$_2$. The decreases in carrier mobility can be attributed to the low-density *O* defects in MoS$_{2-0.19}$O$_{0.19}$, which leads to self-trapping of carriers and random potential fluctuations. Therefore, the low-density *O* defects ultimately result in a decrease in both carrier mobility and on/off ratio.

As $x$ increases to 0.51, the carrier mobility gradually increases to 5.39 cm$^2$ V$^{-1}$ s$^{-1}$, approaching that of MoS$_2$. We attribute this to higher $O$ doping concentration, which leads to a short-range surface defects, thereby tailoring the carrier transport via the NNH of defect-induced localized states in the defective molybdenum disulfide [33].

The NNH can be confirmed by fitting the temperature dependent electric conductance ($G$) following equation:

$$G \propto \exp\left(-\frac{E_a}{k_b T}\right) \tag{1}$$

where $E_a$ is the activation energy for hopping transport in 2D materials, $k_b$ is the Boltzmann constant, and $G$ is the conductance ($I/V_{DS}$) [31]. Fig. 2c illustrates the variation of $I_{DS}$-$V_{GS}$ curves ($V_{DS}$=3V) for MoS$_{2-0.44}$O$_{0.44}$ (i) and MoS$_{2-0.51}$O$_{0.51}$ (ii) as the temperature increases from 310K to 510K. In Fig. 2d, we plot the temperature dependent $G$ at $V_{GS}$=20V and $V_{DS}$=3V. The linear fit of the logarithm of $G$ (MoS$_{2-0.51}$O$_{0.51}$) as a function of $T^{-1}$ provides a clear representation of a strong adherence to Equ. (1) and a conduction channel of the disordered molybdenum disulfide. This suggests that, at high temperatures, the variation in G with T corresponds to thermally activated NNH transport. Additionally, Supplementary III demonstrates the temperature-dependent behavior of $G$ within the range of 220 K and 400 K, thereby confirming the conclusion regarding NNH transport.

We extract $E_a$ from the thermally activated NNH model. Fig. 2e illustrates the relationship between $E_a$ and $V_{GS}$, obtained by fitting the Arrhenius plot of the logarithmic of $G$ dand $T^{-1}$ using Eq. (1). The decreasing $E_a$ values for both MoS$_{2-0.44}$O$_{0.44}$ and MoS$_{2-0.51}$O$_{0.51}$ with increasing $V_{GS}$ suggest the presence of multiple types of $O$ traps [32]. Comparing the $E_a$ values of MoS$_{2-0.44}$O$_{0.44}$ and MoS$_{2-0.51}$O$_{0.51}$, the higher value for MoS$_{2-0.51}$O$_{0.51}$ indicates that a higher concentration of $O$ defects requires more energy for hopping transfer excitation.

**Avalanche multiplication behavior in MoS$_{2-x}$O$_x$ FET**

Fig. 3a demonstrates the forward $I_{DS}$-$V_{DS}$ characteristics of the $MoS_{2-x}O_x$ FET in the high $V_{DS}$ region (0 V < $V_{DS}$ < 40 V) with a fixed $V_{GS}$ of 20 V. To prevent permanent damage to the sample due to repetitive testing, the scanning time for each $I_{DS}$-$V_{DS}$ curve is limited to within 1 minute. The $I_{DS}$ of the $MoS_2$ FET exhibits a sudden increase as $V_{DS}$ increases (Fig. 3a), showing the typical carrier multiplication behavior at high voltages. The $V_{DS}$ value at the point of $I_{DS}$ mutation, known as $V_b$, is measured to be 26.2 V (Fig. 3b). As $V_{GS}$ increases from 20V to 40V, the $V_b$ of the $MoS_2$ FET gradually decreases (Fig. S4d). This contradicts the thermal breakdown variation caused by Joule heating, indicating that the observed carrier multiplication behavior is not a result of Joule heating but rather triggered by impact ionization under high electric fields [8,14]. The impact ionization process under high electric field conditions is illustrated in Fig. S4c. Under the influence of high $V_{DS}$, high-energy electrons (carriers) collide with n-type $MoS_2$, ionizing it and exciting additional electrons. These excited electrons continue to collide and ionize $MoS_2$ under the driving force of high $V_{DS}$, resulting in a chain reaction of avalanche multiplication. In this process, electron collision ionization plays a crucial role, and the impact ionization rate ($\alpha$), as an important parameter for avalanche multiplication, can be determined by Equ. 2.

$$\alpha(V_{DS}) = \frac{1}{n}\frac{dn}{dx} = \frac{1}{L}\left(1 - \frac{1}{M}\right) \tag{2}$$

where $n$ is the electron density [8,14,43,44]; $L$ is the channel length (10 μm); $M$ is the multiplication factor at $V_{DS}$ > $V_b$ presented as $M(V_{DS}) = I_{DS}(V_{DS})/I_{DS}(V_b)$. Based on the above equation, we calculate the variation of $\alpha$ with $V_{DS}$, as shown in Fig. 3c.

In Fig. 3a, the $MoS_{2-0.19}O_{0.19}$, $MoS_{2-0.44}O_{0.44}$, and $MoS_{2-0.51}O_{0.51}$ FETs also exhibit avalanche multiplication phenomenon in the high voltage range, and $V_b$ and $\alpha$ show significant variations with different $x$ values. Fig. 3c illustrates the variation of $\alpha$ with $x$. It is worth noting that $\alpha$ increases with increasing $x$, indicating a gradual increase in the probability of carrier collisions with higher defect concentration. First-principles calculations of the band structure of $MoS_2$- $_x$O$_x$ are performed to further analyze the physical mechanisms of the impact ionization process influenced by $O$ defect concentration in samples. As shown in Fig. 3d-3g, the bandgap of MoS$_{2-x}$O$_x$ gradually decreases with increasing $x$, from 1.67 eV ($x$=0) to 1.58 eV ($x$=0.19), 0.85 eV ($x$=0.44), and 0.55 eV ($x$=0.51). In the impact ionization process, high-energy carriers interact with bound electrons in the valence band through coulomb interactions. Based on the principles of energy and momentum conservation, high-energy carriers transfer energy to the bound electrons, causing them to cross the bandgap into the conduction band, resulting in carrier multiplication. As $x$ increases, the reduction in the bandgap energy of MoS$_{2-x}$O$_x$ facilitates the transfer of excited electrons to the conduction band, thereby enhancing the likelihood of the impact ionization process.

Unlike $α$, the variation of $V_b$ in MoS$_{2-x}$O$_x$ does not show a monotonic change with $x$. Instead, it initially increases to 28.4 V (MoS$_{2-0.19}$O$_{0.19}$) and then gradually decreases to 17.8 V (MoS$_{2-0.44}$O$_{0.44}$) and 12.6 V (MoS$_{2-0.51}$O$_{0.51}$) as shown in Fig. 3b. MoS$_{2-0.19}$O$_{0.19}$ has the highest $V_b$, indicating a difficulty in achieving avalanche multiplication. Although MoS$_{2-0.19}$O$_{0.19}$ has a higher $α$ value compared to MoS$_2$, suggesting a higher probability of collision-induced ionization by charge carriers, its mobility is significantly lower. As a result, the number of charge carriers available for collision ionization in MoS$_{2-0.19}$O$_{0.19}$ is lower than in MoS$_2$ at the same voltage, hindering the occurrence of avalanche multiplication. In contrast, MoS$_{2-0.44}$O$_{0.44}$ and MoS$_{2-0.51}$O$_{0.51}$ rely on the NNH mechanism for electron transport and maintain high mobility even with higher $α$ values. Therefore, MoS$_{2-0.44}$O$_{0.44}$ and MoS$_{2-0.51}$O$_{0.51}$ are more likely to achieve avalanche multiplication.

**Gas response of MoS$_{2-x}$O$_x$ avalanche FET**

To investigate the gas response of MoS$_2$ and MoS$_{2-0.51}$O$_{0.51}$ FETs, we immerse them in a mixture of air and NO$_2$ gas and conduct repeated measurements of avalanche multiplication. With $V_{GS}$ of 20 V, the concentration of NO$_2$ varies from 100 ppb to 250 ppb in increments of 50 ppb. The gas response of the MoS$_2$ FET is shown in Fig. 4a and Fig. 4b, where the output curves ($I_{DS}$-$V_{DS}$) of MoS$_2$ FETs are measured with $V_{DS}$ ranging from 0 to 40 V (in steps of 0.2 V). The output curves exhibit noticeable differences in the avalanche multiplication behavior corresponding to different concentrations of NO$_2$ gas. As the NO$_2$ concentration increases from 100 ppb to 250 ppb, the avalanche multiplication factor $\alpha$ (at $V_{DS}$ = 35 V) decreased from 496.9 cm$^{-1}$ to 151.5 cm$^{-1}$, while $V_b$ increases from 27.3 V to 29.8 V (Fig. 4c). These results indicate that the presence of NO$_2$ gas suppresses avalanche multiplication.

To gain further insights, we performed first-principles calculations to investigate the charge transfer induced by the physisorbed NO$_2$ molecule on the adsorption site of MoS$_2$. The calculations revealed the extraction of electrons induced by the adsorption of NO$_2$ molecules. The NO$_2$ molecule possesses oxidizability due to the presence of unpaired electrons on the nitrogen atoms, which act as electron acceptors. As shown in Fig. 4d, the oxidative NO$_2$ molecule extracts electrons from the n-type MoS$_2$ channel, thereby reducing the carrier concentration in MoS$_2$. This reduced carrier concentration leads to a lower probability of impact ionization, which in turn requires higher electric fields to initiate avalanche multiplication in the NO$_2$-adsorbed MoS$_2$ monolayer.

We calculate the avalanche gas detection response ($R_{AGD}$) following the equation below:

$$R_{AGD}(V_{DS}) = \frac{|R_N(V_{DS}) - R_A(V_{DS})|}{|R_A(V_{DS})|} \times 100\% \qquad (3)$$

where $R_N(V_{DS})$ and $R_A(V_{DS})$ represent the resistance of the MoS$_2$ FET with and without the NO$_2$ adsorption at the voltage of $V_{DS}$ [45,46]. $V_{DS}$ is selected to 35 V above the highest $V_b$ in Fig. 4c, which ensures the occurrence of the avalanche multiplication under different NO$_2$ gas concentrations. Fig. 4e shows an average $R_{AGD}$ of 2.2×10$^2$ %, 1.8×10$^3$ %, 3.6×10$^3$ %, and 4.6×10$^3$ %, corresponding to the NO$_2$ gas of 100 ppb, 150 ppb, 200 ppb, and 250 ppb, respectively. The measurement method of $R_{AGD}$ is described in Supplementary V. The sensitivity of the MoS$_2$ FET, defined as the slope ($S = R_{AGD}/C$) of the resistance against NO$_2$ concentration ($C$), ranges from 2.2% to 18.5% per ppb within a concentration range of 100 ppb to 250 ppb.

The gas performance of MoS$_{2-0.51}$O$_{0.51}$ FET is shown in Fig. 5a and Fig. 5b, which are $I_{DS}$-$V_{DS}$ curves at $V_{GS}$ = 20 V under the different gas atmospheres. When exposed to NO$_2$ gas, the device still exhibits avalanche multiplication behavior with $V_b$ increasing from 13.8 V to 21.8 V along with the gas concentration increasing (from 50 ppb to 200 ppb) and $α$ decreasing from 954.5 cm$^{-1}$ to 669.8 cm$^{-1}$ (Fig. 5c). Meanwhile, $I_{DS}$ of the MoS$_{2-0.51}$O$_{0.51}$ FET decreases for two orders of magnitude in the avalanche region ($V_{DS}$>$V_b$). The variation of $I_{DS}$ is much more significant than the MoS$_2$ FET, demonstrating the MoS$_{2-0.51}$O$_{0.51}$ FET is more sensitive to NO$_2$ gas than the MoS$_2$ FET.

Fig. 5d illustrates the calculated charge transfer of the NO$_2$ molecule on the adsorption site corresponding to the $O$-substituted configuration. Similar to the adsorption on the MoS$_2$ surface, NO$_2$ extracts electrons from the MoS$_{2-0.51}$O$_{0.51}$ surface upon adsorption. As MoS$_{2-0.51}$O$_{0.51}$ is a n-type semiconductor too, its charge carriers remain electrons. When NO$_2$ is adsorbed on the MoS$_{2-0.51}$O$_{0.51}$ surface, it reduces the carrier density of MoS$_{2-0.51}$O$_{0.51}$. Fig. 5e shows the avalanche gas detection response of MoS$_{2-0.51}$O$_{0.51}$ FET at $V_{DS}$ of 35 V. The MoS$_{2-0.51}$O$_{0.51}$ FET has ultra-high $R_{AGD}$ of 5.8×10$^3$ %, 1.5×10$^4$ %, 2.2×10$^4$ %, and 3.3×10$^4$ %, corresponding to NO$_2$ gas of 50 ppb, 100 ppb, 150 ppb, and 200 ppb, respectively. It should be noticed that 50 ppb is the lowest NO$_2$ gas concentration that the mass gas flow controller (MFC) can accurately control in this work and is not the detection limit of the MoS$_{2-0.51}$O$_{0.51}$ FET. Supplementary VI displays the avalanche behavior of the MoS$_{2-0.51}$O$_{0.51}$ FET with the continuous decrease of the NO$_2$ concentration from 50 ppb over time, and distinct output curves are still observed at concentrations less than 50 ppb, indicating that the MoS$_{2-0.51}$O$_{0.51}$ FET exhibits ultra-sensitivity to NO$_2$ molecules.

The limit of detection (LOD) is a crucial performance metric for gas sensors, representing the concentration of the analyte that produces a response three times higher than the noise level of the device (LOD=3×noise/sensitivity). Supplementary V displays the noise data of the measurement of 100 ppb $NO_2$. An root-mean-square value of $\Delta R_{AGD}/R_{AGD}$ noise of approximately 0.7% is obtained, leading to a calculated LOD of approximately $3\times0.7\%/(1.5\times10^4\%/ppb)=1.4\times10^{-4}$ ppb. Compared to resistive-type gas sensors, the avalanche-type gas sensor exhibits significantly higher sensitivity despite their higher noise levels, resulting in LOD values much lower than those of resistive-type ones.

Supplementary VII compares the gas-sensing response of our Avalanche-type sensors with reported resistance-type ones based on TMDCs, where the value of $R_{AGD}$ is characterized consistently following Eq. (3). We divide TMDCs-based gas sensors into homostructure (green circle) and heterostructure (blue circle) according to the composition of materials. As one can see, the avalanche-type $MoS_{2-x}O_x$ FETs exhibit more significant gas-sensing performance than reported resistance-type sensors. In particular, the $R_{AGD}$ of $MoS_{2-x}O_x$ FETs is two orders of magnitude greater than resistance-type ones, demonstrating that gas-sensing performances can be significantly improved by taking advantage of the avalanche multiplication.

Please note, the avalanche-type $MoS_2$ (or $MoS_{2-x}O_x$) FET has the same structural characteristic as the homostructure gas sensor except for the gas-detecting method (see Experimental Methods and Fig. S5). Traditional resistor-type detection typically performs measurements at a constant low operating voltage ($V_{DS} < 10$ V), and the gas adsorption (or desorption) modifies the current (or resistance) curve. In comparison, the avalanche-type one operates under a high impulse voltage ($V_{DS} > 20$ V) keeping itself in the avalanche regime, and the adsorbed (or desorbed) gas molecule disturbs the avalanche multiplication state intriguing a more drastic change in the current curve. Although the structure and materials of avalanche-type gas sensors are the same as the resistor-type ones, the gas-sensing performance of the avalanche-type TMDCs-based FET is greatly enhanced, demonstrating the advantage of the mechanism of avalanche multiplication.

**Conclusion**

Our work demonstrates the potential of using NNH to reduce the breakdown voltage of $MoS_{2-x}O_x$ FETs and enable ultrasensitive gas detection. By substituting *O* atoms for *S* atoms in a monolayer of $MoS_2$, we make $MoS_{2-x}O_x$ with carefully controlled *O* doping levels. The high level of oxygen doping in $MoS_{2-x}O_x$ (x > 0.41) exhibit a NNH behavior, significantly enhancing electron mobility. Meanwhile, the *O* doping not only reduces the bandgap of TMDCs but also enhances ion collision rates. These improvement decreases the breakdown voltage of avalanche multiplication. We also propose avalanche multiplication in $MoS_{2-x}O_x$ is the promising option for ppb-level gas sensing, exhibiting a response of $5.8 \times 10^3$ % to $NO_2$ gas at 50 ppb concentration at room temperature, two orders higher than the resistance-type gas detectors based on TMDCs. This work highlights the effectiveness of hopping transfer induced by high-density oxygen defects in reducing the breakdown voltage of $MoS_{2-x}O_x$ FETs and enabling ultrasensitive gas detection, and opens up new possibilities for the development of highly sensitive gas sensors.

**Experimental Methods**

*Sample preparation*

The MoS$_2$ monolayers were grown on a SiO$_2$/Si substrate via the CVD method. After then, the MoS$_2$ was transferred to a SiO$_2$/Si substrate with prefabricated Au electrodes (Au: 50 nm, SiO$_2$: 300 nm) via a standard PDMS-Assisted Transfer Process [47]. A thin layer of PDMS was placed on top of MoS$_2$ on SiO$_2$/Si, the assembly was immersed into the 90 °C 2-mol/L KOH solution for 15 minutes to etch SiO$_2$ away. The sample was then taken out and rinsed three times with DI water. PDMS was peeled off from SiO$_2$ and adhered to the slide. Next, the PDMS was stamped on the target substrate via the transfer stage and thermally released at 80 °C. Finally, the assembled device was annealed in an argon-protected tube furnace at 200 °C.

*Characterization*

The fabricated devices were characterized using an optical microscope (Leica DM2700), and AFM (Bruker, Dimension Icon). The electrical characteristics were measured via Keithley 4200A-SCS at room temperature. The power of the laser used in Raman spectra measurements was 1mW, and the wavelength was 532 nm.

*Gas-detecting Method*

The gas sensing measurement was carried out in a gas chamber with a quartz transparent window. The concentration of the NO$_2$ gas was controlled by two mass flow controllers (S49 32/MT). During the measurement, $V_{DS}$ of a periodic sawtooth signal was applied to the FETs with a frequency of $8.3 \times 10^{-3}$ Hz, and $V_{GS}$ was constant at a specific value. When the gas was off, the FET was illuminated by white light (408 lx, 0.9 mW/cm$^2$) for 1 min to promote the desorption of gas molecules. Supplementary V provides more detailed information on the measurement method.

**Supporting Information**.

Here, you will find the detailed information for " Characterization of monolayer MoS$_2$ ", " The output curve of monolayer MoS$_2$ FET in low $V_{DS}$ regime ", " The NNH fit of the logarithm of $G$ of MoS$_{2-0.51}$O$_{0.51}$ ", " The avalanche multiplication behavior of pristine MoS$_2$ FET ", " The detail of measurement method for $R_{ADR}$ ", " The output curve of the MoS$_{2-0.51}$O$_{0.51}$ FET with the continuous decrease of the NO$_2$ concentration from 50 ppb over time ", " Comparing of the gas-sensing response based on TMDCs ".

## ACKNOWLEDGMENT

This work is supported by the National Natural Science Foundation of China (No. 12122508).

REFERENCES

[1] D. J. Late, Y. K. Huang, B. Liu, J. Acharya, S. N. Shirodkar, J. Luo, A. Yan, D. Charles, U. V. Waghmare, V. P. Dravid, C. N. Rao, Sensing behavior of atomically thin-layered MoS$_2$ transistors, *ACS Nano 7* (6) (2013) 4879-4891. https://doi.org/10.1021/nn400026u.

[2] C. Xie, C. Mak, X. Tao, F. Yan, Photodetectors Based on Two-Dimensional Layered Materials Beyond Graphene, Adv. Funct. Mater. 27 (19) (2017) 1603886. https://doi.org/10.1002/adfm.201603886.

[3] F. H. Koppens, T. Mueller, P. Avouris, A. C. Ferrari, M. S. Vitiello, M. Polini, Photodetectors based on graphene, other two-dimensional materials and hybrid systems, Nat. Nanotechnol. 9 (10) (2014) 780-793. https://doi.org/10.1038/nnano.2014.215.

[4] X. Liu, T. Ma, N. Pinna, J. Zhang, Two-Dimensional Nanostructured Materials for Gas Sensing, Adv. Funct. Mater. 27 (37) (2017) 1702168. https://doi.org/10.1002/adfm.201702168.

[5] Q. H. Wang, K. Kalantar-Zadeh, A. Kis, J. N. Coleman, M. S. Strano, Electronics and optoelectronics of two-dimensional transition metal dichalcogenides, Nat. Nanotechnol. 7 (11) (2012) 699-712. https://doi.org/10.1038/nnano.2012.193.

[6] D. Jariwala, V. K. Sangwan, L. J. Lauhon, T. J. Marks, M. C. Hersam, Emerging device applications for semiconducting two-dimensional transition metal dichalcogenides, ACS Nano 8 (2) (2014) 1102-1120. https://doi.org/10.1021/nn500064s.

[7] S. J. Kim, K. Choi, B. Lee, Y. Kim, B. H. Hong, Materials for Flexible, Stretchable Electronics: Graphene and 2D Materials, Annu. Rev. Mater. Sci. 45 (1) (2015) 63-84. ! https://doi.org/10.1146/annurev-matsci-070214-020901.

[8] J. Pak, Y. Jang, J. Byun, K. Cho, T. Y. Kim, J. K. Kim, B. Y. Choi, J. Shin, Y. Hong, S. Chung, T. Lee, Two-Dimensional Thickness-Dependent Avalanche Breakdown Phenomena in MoS$_2$ Field-Effect Transistors under High Electric Fields, ACS Nano 12 (7) (2018) 7109-7116. ! https://doi.org/10.1021/acsnano.8b02925.

[9] J. Kim, K. Cho, J. Pak, W. Lee, J. Seo, J. K. Kim, J. Shin, J. Jang, K. Y. Baek, J. Lee, S. Chung, K. Kang, T. Lee, Channel-Length-Modulated Avalanche Multiplication in Ambipolar WSe$_2$ Field-Effect Transistors, ACS Nano 16 (4) (2022) 5376-5383. https://doi.org/10.1021/acsnano.1c08104.

[10] W. Deng, X. Chen, Y. Li, C. You, F. Chu, S. Li, B. An, Y. Ma, L. Liao, Y. Zhang, Strain Effect

Enhanced Ultrasensitive MoS$_2$ Nanoscroll Avalanche Photodetector, J. Phys. Chem. Lett. 11 (11) (2020) 4490-4497. https://doi.org/10.1021/acs.jpclett.0c00861.

[11] L. Y. Meng, N. N. Zhang, M. L. Yang, X. X. Yuan, M. L. Liu, H. Y. Hu, L. M. Wang, Low-voltage and high-gain WSe$_2$ avalanche phototransistor with an out-of-plane WSe$_2$/WS$_2$ heterojunction, Nano Res. 16 (2022) 3422. https://doi.org/10.1007/s12274-022-4954-6.

[12] H. Choi, S. Choi, T. Kang, H. Son, C. Kang, E. Hwang, S. Lee, Broad‐Spectrum Photodetection with High Sensitivity Via Avalanche Multiplication in WSe$_2$, Adv. Opt. Mater. 10 (22) (2022) 2201196. https://doi.org/ 10.1002/adom.202201196.

[13] B. Son, Y. Wang, M. Luo, K. Lu, Y. Kim, H. J. Joo, Y. Yi, C. Wang, Q. J. Wang, S. H. Chae, D. Nam, Efficient Avalanche Photodiodes with a WSe$_2$/MoS$_2$ Heterostructure via Two-Photon Absorption, Nano Lett. 22 (23) (2022) 9516-9522. https://doi.org/ 10.1021/acs.nanolett.2c03629.

[14] J. Seo, J. H. Lee, J. Pak, K. Cho, J. K. Kim, J. Kim, J. Jang, H. Ahn, S. C. Lim, S. Chung, K. Kang, T. Lee, Ultrasensitive Photodetection in MoS$_2$ Avalanche Phototransistors, Adv. Sci. 8 (19) (2021) 2102437. https://doi.org/10.1002/advs.202102437.

[15] F. Ahmed, Y. D. Kim, Z. Yang, P. He, E. Hwang, H. Yang, J. Hone, W. J. Yoo, Impact ionization by hot carriers in a black phosphorus field effect transistor, Nat. Commun. 9 (1) (2018) 3414. https://doi.org/10.1038/s41467-018-05981-0.

[16] A. Gao, J. Lai, Y. Wang, Z. Zhu, J. Zeng, G. Yu, N. Wang, W. Chen, T. Cao, W. Hu, D. Sun, X. Chen, F. Miao, Y. Shi, X. Wang, Observation of ballistic avalanche phenomena in nanoscale vertical InSe/BP heterostructures, Nat. Nanotechnol. 14 (3) (2019) 217-222. https://doi.org/10.1038/s41565-018-0348-z.

[17] Y. Hattori, T. Taniguchi, K. Watanabe, K. Nagashio, Impact ionization and transport properties of hexagonal boron nitride in a constant-voltage measurement, Phys. Rev. B 97 (4) (2018) 045425. https://doi.org/10.1103/PhysRevB.97.045425.

[18] Z. Zhang, B. Cheng, J. Lim, A. Gao, L. Lyu, T. Cao, S. Wang, Z. A. Li, Q. Wu, L. K. Ang, Y. S. Ang, S. J. Liang, F. Miao, Approaching the Intrinsic Threshold Breakdown Voltage and Ultrahigh Gain in a Graphite/InSe Schottky Photodetector, Adv. Mater. 34 (47) (2022) 2206196. https://doi.org/10.1002/adma.202206196.

[19] V. K. Sangwan, J. Kang, D. Lam, J. T. Gish, S. A. Wells, J. Luxa, J. P. Male, G. J. Snyder, Z. Sofer, M.

C. Intrinsic carrier multiplication in layered $Bi_2O_2Se$ avalanche photodiodes with gain bandwidth product exceeding 1 GHz, Nano Res. 14 (6) (2020) 1961-1966. https://doi.org/10.1007/s12274-020-3059-3.

[20] S. Lei, F. Wen, L. Ge, S. Najmaei, A. George, Y. Gong, W. Gao, Z. Jin, B. Li, J. Lou, J. Kono, R. Vajtai, P. Ajayan, N. J. Halas, An Atomically Layered InSe Avalanche Photodetector, Nano Lett. 15 (5) (2015) 3048-3055. https://doi.org/10.1021/acs.nanolett.5b00016.

[21] J. Jia, J. Jeon, J. H. Park, B. H. Lee, E. Hwang, S. Lee, Avalanche Carrier Multiplication in Multilayer Black Phosphorus and Avalanche Photodetector, Small 15 (38) (2019) 1805352. https://doi.org/10.1002/smll.201805352.

[22] S. Das Sarma, S. Adam, E. H. Hwang, E. Rossi, Electronic transport in two-dimensional graphene, Rev. Mod. Phys. 83 (2) (2011) 407-470. https://doi.org/10.1103/RevModPhys.83.407.

[23] S. Kc, R. C. Longo, R. Addou, R. M. Wallace, K. Cho, Impact of intrinsic atomic defects on the electronic structure of $MoS_2$ monolayers, Nanotechnology 25 (37) (2014) 375703. https://doi.org/10.1088/0957-4484/25/37/375703.

[24] S. Yuan, R. Roldán, M. I. Katsnelson, F. Guinea, Effect of point defects on the optical and transport properties of $MoS_2$ and $WS_2$, Phys. Rev. B 90 (4) (2014) 041402. https://doi.org/10.1103/PhysRevB.90.041402.

[25] H. Komsa, J. Kotakoski, S. Kurasch, O. Lehtinen, U. Kaiser, A. V. Krasheninnikov, Two-Dimensional Transition Metal Dichalcogenides under Electron Irradiation: Defect Production and Doping, Phys. Rev. Lett. 109 (3) (2012) 035503. https://doi.org/10.1103/PhysRevLett.109.035503.

[26] X. Zhang, L. Gao, H. Yu, Q. Liao, Z. Kang, Z. Zhang, Y. Zhang, Single-Atom Vacancy Doping in Two-Dimensional Transition Metal Dichalcogenides, Acc. Mater. Res. 2 (8) (2021) 655-668. https://doi.org/10.1021/accountsmr.1c00097.

[27] Y. Liu, Z. Gao, Y. Tan, F. Chen, Enhancement of Out-of-Plane Charge Transport in a Vertically Stacked Two-Dimensional Heterostructure Using Point Defects, ACS Nano 12 (10) (2018) 10529-10536. https://doi.org/10.1021/acsnano.8b06503.

[28] Y. J. Zheng, Y. Chen, Y. L. Huang, P. K. Gogoi, M.-Y. Li, L.-J. Li, P. E. Trevisanutto, Q. Wang, S. J. Pennycook, A. T. S. Wee, S. Y. Quek, Point Defects and Localized Excitons in 2D $WSe_2$. ACS Nano 13 (5) (2019) 6050-6059. https://doi.org/10.1021/acsnano.9b02316.

[29] Z. Yu, Z. Y. Ong, S. Li, J. B. Xu, G. Zhang, Y. W. Zhang, Y. Shi, X. Wang, Analyzing the Carrier Mobility in Transition-Metal Dichalcogenide MoS$_2$ Field-Effect Transistors, Adv. Funct. Mater. 27 (19) (2017) 1604093. https://doi.org/10.1002/adfm.201604093.

[30] L. Li, M.-F. Lin, X. Zhang, A. Britz, A. Krishnamoorthy, R. Ma, R. K. Kalia, A. Nakano, P. Vashishta, P. Ajayan, M. C. Hoffmann, D. M. Fritz, U. Bergmann, O. V. Prezhdo, Phonon-Suppressed Auger Scattering of Charge Carriers in Defective Two-Dimensional Transition Metal Dichalcogenides, Nano Lett. 19 (9) (2019) 6078-6086. https://doi.org/10.1021/acs.nanolett.9b02005.

[31] M. G. Stanford, P. R. Pudasaini, E. T. Gallmeier, N. Cross, L. Liang, A. Oyedele, G. Duscher, M. Mahjouri-Samani, K. Wang, K. Xiao, D. B. Geohegan, A. Belianinov, B. G. Sumpter, P. D. Rack, High Conduction Hopping Behavior Induced in Transition Metal Dichalcogenides by Percolating Defect Networks: Toward Atomically Thin Circuits, Adv. Funct. Mater. 27 (36) (2017) 1702829. https://doi.org/10.1002/adfm.201702829

[32] F. Ahmed, A. M. Shafi, D. M. A. Mackenzie, M. A. Qureshi, H. A. Fernandez, H. H. Yoon, M. G. Uddin, M. Kuittinen, Z. Sun, H. Lipsanen, Multilayer MoTe$_2$ Field-Effect Transistor at High Temperatures, Adv. Mater. Interfaces 8 (22) (2021) 2100950. https://doi.org/10.1002/admi.202100950

[33] H. Qiu, T. Xu, Z. Wang, W. Ren, H. Nan, Z. Ni, Q. Chen, S. Yuan, F. Miao, F. Song, G. Long, Y. Shi, L. Sun, J. Wang, X. Wang, Hopping transport through defect-induced localized states in molybdenum disulphide, Nat. Commun. 4 (1) (2013) 2642. https://doi.org/10.1038/ncomms3642

[34] M. Y. Han, J. C. Brant, P. Kim, Electron Transport in Disordered Graphene Nanoribbons, Phys. Rev. Lett. 104 (5) (2010) 056801. https://doi.org/10.1103/PhysRevB.88.205414

[35] J. Y. Kim, I. Kaganovich, H.-C. Lee, Review of the gas breakdown physics and nanomaterial-based ionization gas sensors and their applications, Plasma Sources Sci Technol. 31 (3) (2022) 033001. https://doi.org/10.1088/1361-6595/ac4574

[36] S. Bakhshi Sichani, A. Nikfarjam, H. Hajghassem, A novel miniature planar gas ionization sensor based on selective growth of ZnO nanowires, Sens. Actuator A Phys. 288 (2019) 55-60. https://doi.org/10.1016/j.sna.2019.01.024.

[37] R. Mohammadpour, H. Ahmadvand, A. Iraji zad, A novel field ionization gas sensor based on self-organized CuO nanowire arrays, Sens. Actuator A Phys. 216 (2014) 202-206. https://doi.org/10.1016/j.sna.2014.04.038.

[38] J. Zhang, Y. Zhang, Z. Pan, S. Yang, J. Shi, S. Li, D. Min, X. Li, X. Wang, D. Liu, A. Yang, Properties of a weakly ionized NO gas sensor based on multi-walled carbon nanotubes, Appl. Phys. Lett. 107 (9) (2015) 093104. https://doi.org/10.1063/1.4930020

[39] P. Abedini Sohi, M. Kahrizi, Low-Voltage Gas Field Ionization Tunneling Sensor Using Silicon Nanostructures, IEEE Sens. J. 18 (15) (2018) 6092-6096. https://doi.org/10.1109/JSEN.2018.2846254.

[40] S. Bertolazzi, S. Bonacchi, G. Nan, A. Pershin, D. Beljonne, P. Samorì, Engineering Chemically Active Defects in Monolayer $MoS_2$ Transistors via Ion-Beam Irradiation and Their Healing via Vapor Deposition of Alkanethiols, Adv. Mater. 29 (18) (2017) 1606760. ! https://doi.org/10.1002/adma.201606760.

[41] X. Xiong, F. Wu, Y. Ouyang, Y. Liu, Z. Wang, H. Tian, M. Dong, Oxygen Incorporated $MoS_2$ for Rectification-Mediated Resistive Switching and Artificial Neural Network, Adv. Funct. Mater. (2023) 2213348. https://doi.org/10.1002/adfm.202213348.

[42] D. M. Sim, M. Kim, S. Yim, M.-J. Choi, J. Choi, S. Yoo, Y. S. Jung, Controlled Doping of Vacancy-Containing Few-Layer $MoS_2$ via Highly Stable Thiol-Based Molecular Chemisorption, ACS Nano 9 (12) (2015) 12115-12123. https://doi.org/10.1021/acsnano.5b05173

[43] A. G. Chynoweth, Uniform Silicon p-n Junctions. II. Ionization Rates for Electrons, J. Appl. Phys. 31 (7) (1960) 1161-1165. https://doi.org/10.1063/1.1735795.

[44] K. G. McKay, K. B. McAfee, Electron Multiplication in Silicon and Germanium, Phys. Rev. 91 (5) (1953) 1079-1084. https://doi.org/10.1103/PhysRev.91.1079.

[45] B. Cho, M. G. Hahm, M. Choi, J. Yoon, A. R. Kim, Y. J. Lee, S. G. Park, J. D. Kwon, C. S. Kim, M. Song, Y. Jeong, K. S. Nam, S. Lee, T. J. Yoo, C. G. Kang, B. H. Lee, H. C. Ko, P. M. Ajayan, D. H. Kim, Charge-transfer-based gas sensing using atomic-layer $MoS_2$, Sci. Rep. 5 (2015) 8052. ! https://doi.org/10.1038/srep08052.

[46] S. Y. Cho, S. J. Kim, Y. Lee, J. S. Kim, W. B. Jung, H. W. Yoo, J. Kim, H. T. Jung, Highly Enhanced Gas Adsorption Properties in Vertically Aligned $MoS_2$ Layers, ACS Nano 9 (9) (2015) 9314-9321. ! https://doi.org/10.1021/acsnano.5b04504.

[47] D. B. Trivedi, G. Turgut, Y. Qin, M. Y. Sayyad, D. Hajra, M. Howell, L. Liu, S. Yang, N. H. Patoary, H. Li, M. M. Petric, M. Meyer, M. Kremser, M. Barbone, G. Soavi, A. V. Stier, K. Muller, S. Yang, I. S. Esqueda, H. Zhuang, J. J. Finley, S. Tongay, Room-Temperature Synthesis of 2D Janus Crystals

and their Heterostructures, Adv. Mater. 32 (50) (2020) 2006320. !

https://doi.org/10.1002/adma.202006320.

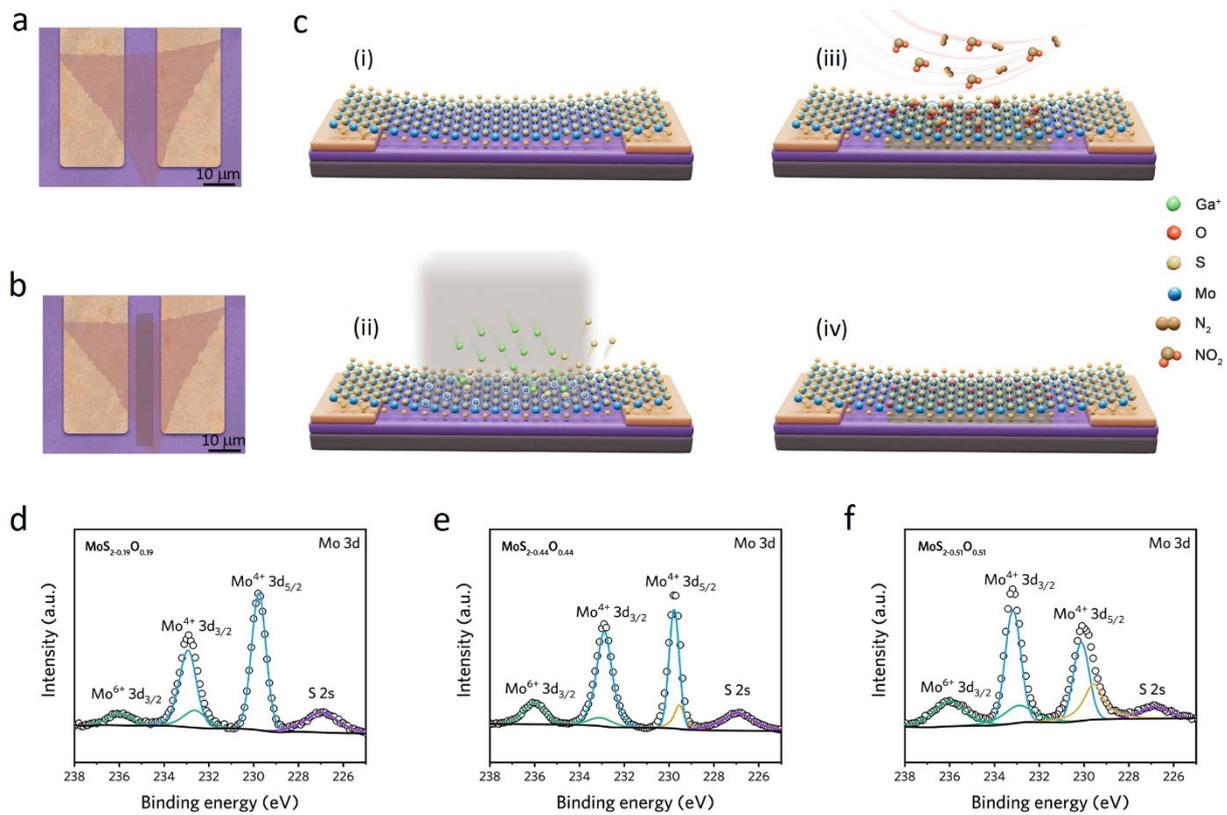

**Fig. 1.** Preparation of monolayer $MoS_{2-x}O_x$. SEM images of the molybdic sulfide before (**a**) and after (**b**) ion irradiation. (**c**) Preparation process of the $MoS_{2-x}O_x$. XPS spectra of $MoS_{2-0.19}O_{0.19}$ (**d**), $MoS_{2-0.44}O_{0.44}$ (**e**), and $MoS_{2-0.51}O_{0.51}$ (**f**).

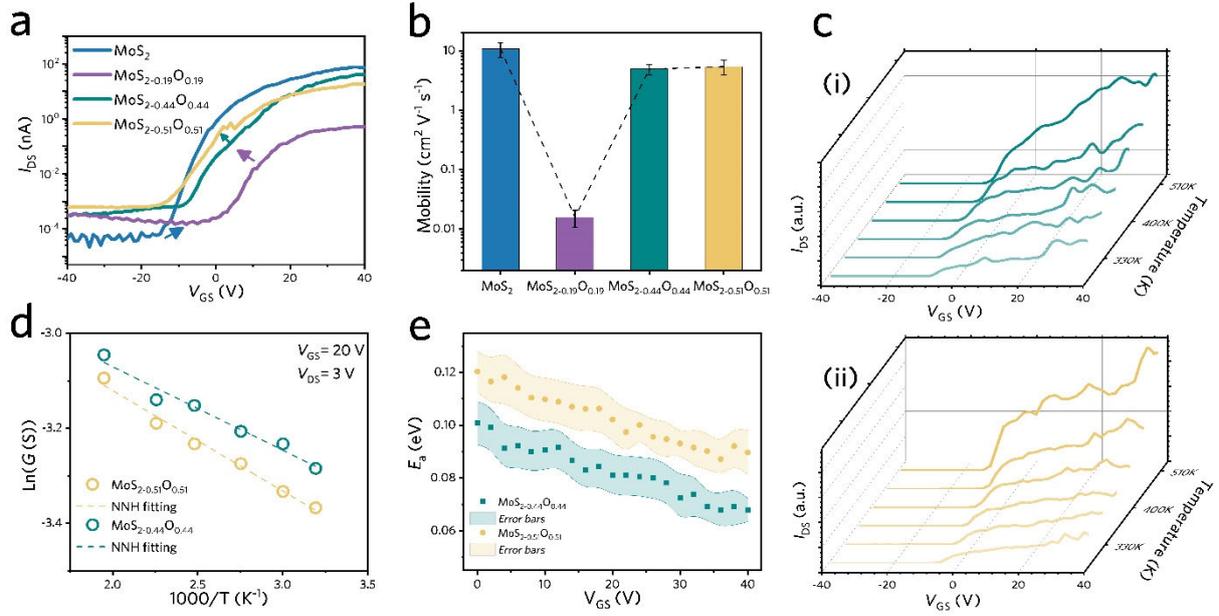

**Fig. 2.** NNH transfer behavior in monolayer $MoS_{2-x}O_x$. (**a**) Transfer characteristics of $MoS_2$ and $MoS_{2-x}O_x$ ($x = 0.19$, 0.44 and 0.51) at room temperature, presented in the logarithmic scale. (**b**) Mobility variation of $MoS_{2-x}O_x$ corresponding to $x$. (**c**) variation of $I_{DS}$-$V_{GS}$ curves ($V_{DS}$=3V) for $MoS_{2-0.44}O_{0.44}$ (i) and $MoS_{2-0.51}O_{0.51}$ (ii) as the temperature increases from 310 K to 510 K. (**d**) The linear fit of the logarithm of $G$ as a function of $T^{-1}$. (**e**) Relationship between $E_a$ and $V_{GS}$.

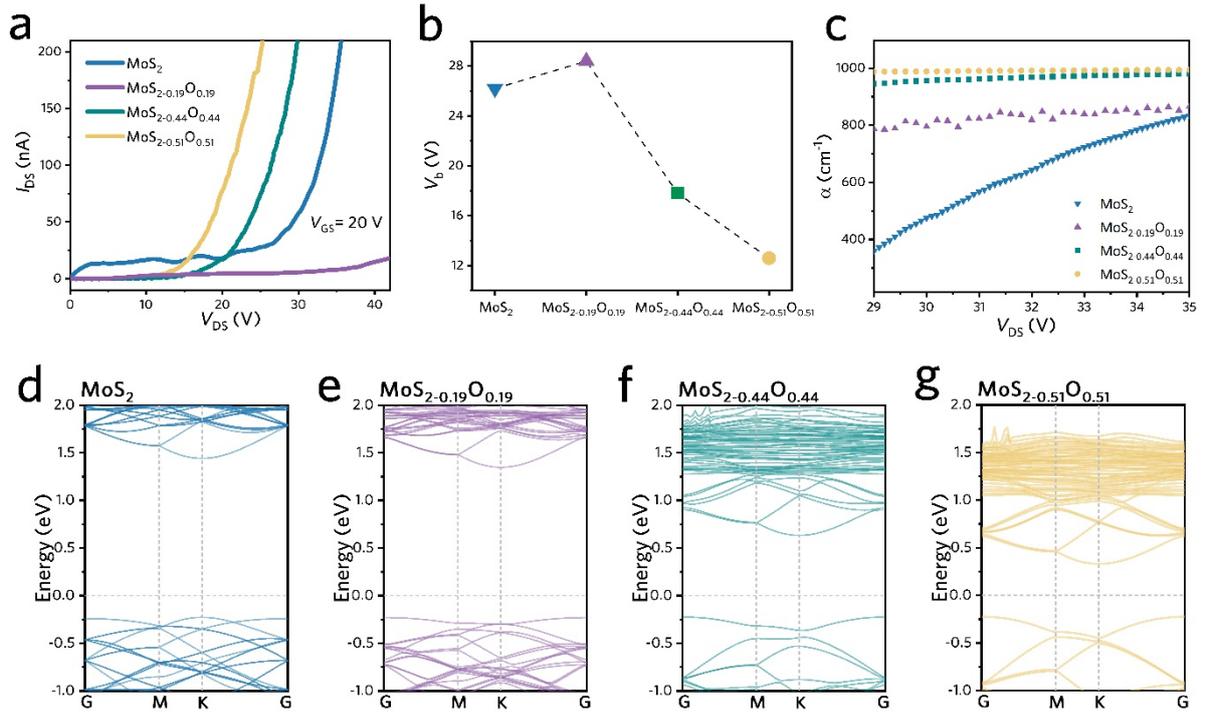

**Fig. 3.** Avalanche characterization of monolayer MoS$_{2-x}$O$_x$. (**a**) $I_{DS}$-$V_{DS}$ curves of MoS$_{2-x}$O$_x$ FET with a fixed $V_{GS}$ of 20 V. (**b**) $V_b$ of MoS$_{2-x}$O$_x$. (**c**) Variation of $\alpha$ with $V_{DS}$. Electronic structures of MoS$_2$ (**d**), MoS$_{2-0.19}$O$_{0.19}$ (**e**), MoS$_{2-0.44}$O$_{0.44}$ (**f**), and MoS$_{2-0.51}$O$_{0.51}$ (**g**).

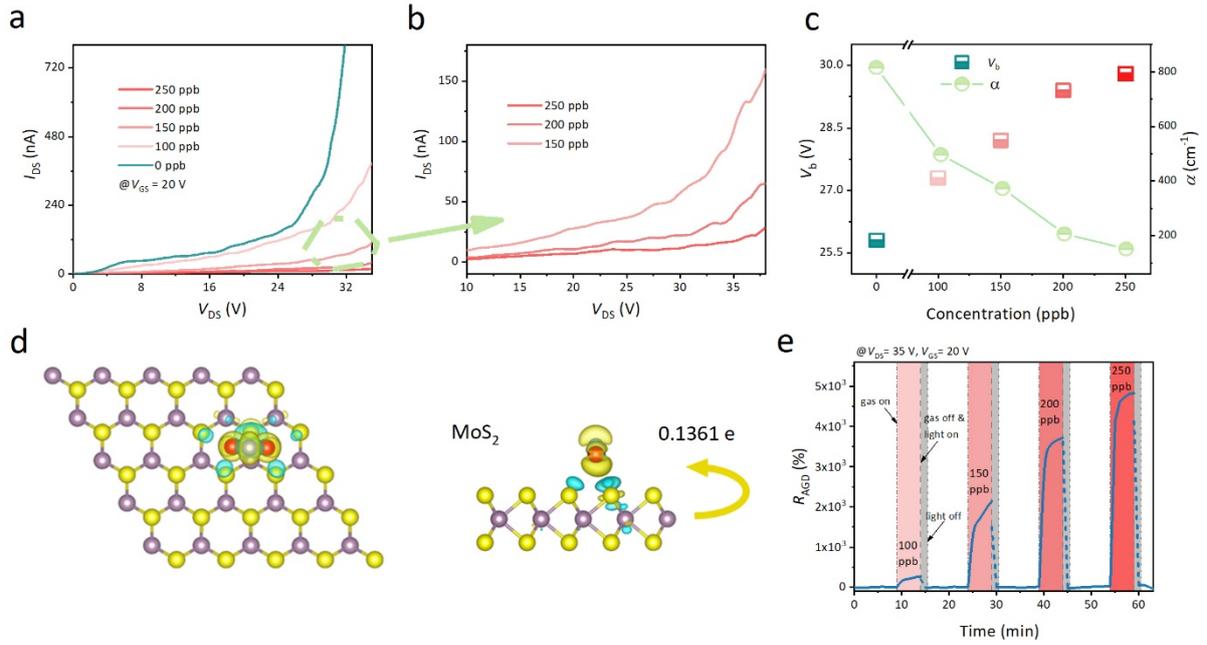

**Fig. 4.** Gas response of MoS$_2$ FET with NO$_2$ adsorption. (**a**) $I_{DS}$-$V_{DS}$ curves of a monolayer MoS$_2$ FET at $V_{GS}$ = 20 V in ambient air and various concentrations of NO$_2$ ranging from 100 ppb to 250 ppb with a step of 50 ppb. (**b**) A larger view of the potion in green box shown in (**a**). (**c**) $V_b$ and $\alpha$ ($V_{DS}$ = 35 V) values at $V_{GS}$ = 20 V in different gas atmospheres, represented as balls and squares, respectively. (**d**) Calculated charge transfer of MoS$_2$ FET with NO$_2$ adsorption. (**e**) Measurements of avalanche gas detection response ($R_{ADR}$) in different gas atmospheres, following the one described in Section *Gas-detecting method*.

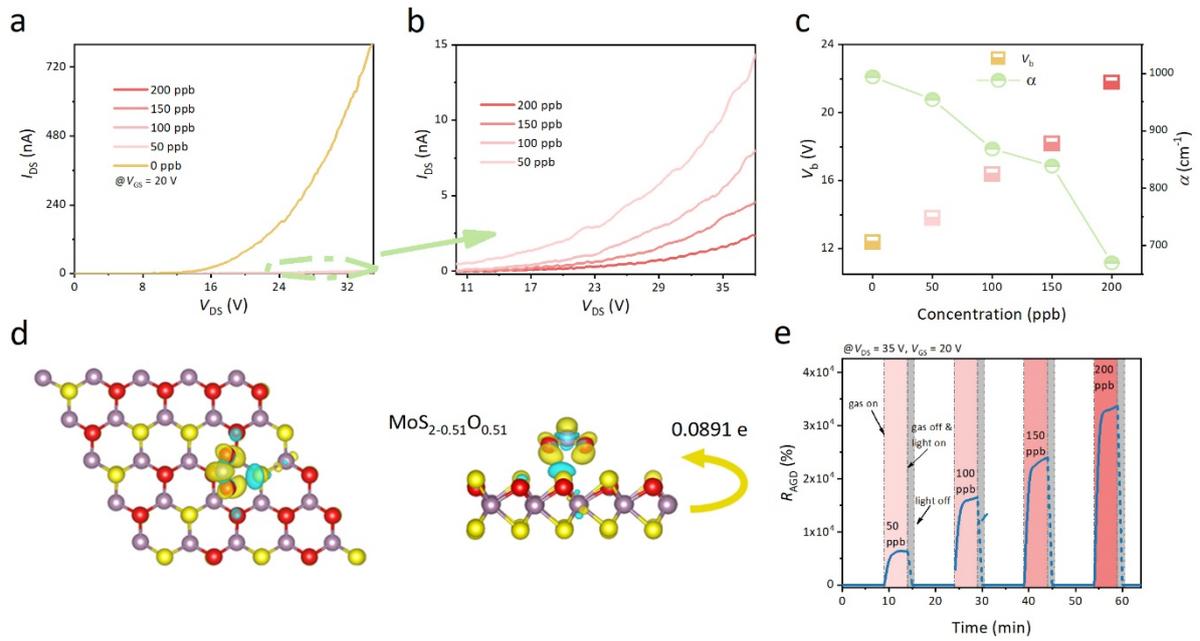

**Fig. 5.** Gas response of $MoS_{2-0.51}O_{0.51}$ FET with $NO_2$ adsorption. (**a**) $I_{DS}$-$V_{DS}$ curves of $MoS_{2-0.51}O_{0.51}$ FET at $V_{GS}$ = 20 V in ambient air and various concentrations of $NO_2$ ranging from 50 ppb to 200 ppb with a step of 50 ppb. (**b**) A larger view of the potion in green box shown in (**a**). (**c**) $V_b$ and $α$ ($V_{DS}$ = 35 V) values at $V_{GS}$ = 20 V in different gas atmospheres, respectively. (**d**) Calculated charge transfer of $MoS_{2-0.51}O_{0.51}$ FET with $NO_2$ adsorption. (**e**) Measurements of avalanche gas detection response ($R_{ADR}$) in different gas atmospheres, following the one described in Section *Gas-detecting method*.

26